\documentclass[11pt]{article}
\usepackage{amsmath, amssymb, geometry, setspace}
\usepackage{float}
\usepackage{caption}
\usepackage{booktabs}
\usepackage{array}
\newcolumntype{L}[1]{>{\raggedright\arraybackslash}p{#1}}
\usepackage{parskip}
\usepackage{graphicx}
\usepackage[colorlinks=true, linkcolor=blue, citecolor=blue, urlcolor=blue]{hyperref}
\usepackage{natbib}
\newcommand{\Var}{\mathrm{Var}}

\newcommand{\E}{\mathbb{E}}
\newcommand{\SE}{\mathrm{SE}}
\newcommand{\nbar}{\bar n}
\newcommand{\cv}{cv}
\newcommand{\vmac}{\sigma^2_{\mathrm{macro}}}
\newcommand{\vres}{\sigma^2_{\mathrm{res}}}
\newcommand{\Smacro}{S_{macro}}
\newcommand{\Scl}{S_{cl}}
\newcommand{\Stime}{S_{time}}
\newcommand{\Sint}{S_{int}}
\newcommand{\Sres}{S_{res}}
\newcommand{\VRmax}{\mathrm{VR}_{\max}}
\newcommand{\VRatt}{\mathrm{VR}_{\mathrm{att}}}

\title{CUPED on Steroids: Multivariate Covariate Adjustment for Switchback Experiments}
\author{Sergei Pankratev and Palash Arora}
\date{}

\begin{document}
\maketitle

\begin{abstract}
Controlled-experiment Using Pre-Experiment Data (CUPED) reduces the variance of the treatment effect estimator in online experiments by adjusting the in-experiment outcome metric using its lagged pre-experiment value.
This method can be strengthened by enriching its covariate set while keeping it automatable and guarding against overfitting and covariate leakage.
In switchback and related clustered experiment designs, this enrichment may be especially fruitful since randomization-unit-level variation in the outcome amplifies the variance of the treatment effect estimator, so covariates that explain it yield disproportionately large efficiency gains.
We develop an extended CUPED framework for switchback and related clustered designs that introduces sizeable improvement in variance reduction efficiency relative to conventional CUPED while remaining lightweight and free of assumptions beyond those already implicit in CUPED.
The framework combines multiple historical lags with cyclic hour-of-day encodings and cluster- and hour-level fixed-effect dummies in the pre-period prediction model---covariates that target randomization-unit-level outcome variation---and uses cross-fitting so that realized variance reduction is not inflated by in-sample overfitting.
In simulation, we analyze how each component affects validity and efficiency of this method and show that the full framework yields large gains in statistical power.
We further validate the approach on publicly available trip records from the New York City Taxi and Limousine Commission.
Finally, we derive a closed-form ceiling on the variance reduction attainable by any covariate adjustment and connect our empirical findings to the composition-dependent limits this ceiling imposes.
\end{abstract}

\section{Introduction}
\label{sec:intro}

Variance reduction is central to the practice of online experimentation, because it improves sensitivity without requiring additional sample size or longer experiment duration \citep{kohavi2020trustworthy}.
The canonical technique, Controlled-experiment Using Pre-Experiment Data (CUPED), adjusts the outcome metric using its pre-experiment value, exploiting the correlation between a unit's pre-period and in-period behavior to remove predictable variation \citep{deng2013improving}.\footnote{
    Throughout the paper, \emph{outcome} denotes the business metric whose treatment effect the experiment is designed to estimate---for example, trip duration, conversion rate, or order value.
}
In its most common form, CUPED relies on a single covariate, the lagged metric, and the achievable reduction is therefore bounded by the share of the current-period outcome that one pre-period value alone can explain.
A natural way to strengthen the adjustment is to enlarge the information used to predict the outcome.
At one extreme, machine-learning-based extensions replace the single lagged covariate with the prediction of a flexible model trained on many pre-experiment attributes, a strategy often referred to as Control Using Prediction As Covariate (CUPAC) \citep{poyarkov2016boosted}.
Enriching the covariate set with multiple historical lags and structured fixed effects sits between these two poles: it extends CUPED beyond a single-lag adjustment while stopping short of a fully flexible CUPAC model, requires no change to the experimental design, and remains lightweight, automatable, and easy to integrate into existing analysis pipelines.

Switchback experiments, in which treatment is assigned at the level of a spatial cluster crossed with a time period, are widely used in marketplace and operational experimentation because randomizing individual units would allow treatment effects to propagate through network interference or shared local conditions \citep{bojinov2023design}.
In such settings, cell-level variation in the outcome---shocks shared by every observation within the same randomization unit---accounts for most of the uncertainty in the treatment-effect estimate, and cluster size imbalance amplifies that contribution \citep{pankratev2026powerful}.
Because panel datasets from switchback experiments are typically unbalanced---some cluster-by-time cells contain far more observations than others---CUPED is almost always applied after aggregating pre-period outcome values to the cell, the unit of randomization.
The adjustment therefore targets cell-level variation in the outcome rather than variation specific to individual units, making it well suited to reducing the portion of uncertainty in the estimate that imbalance inflates most severely.\footnote{
    We adopt the terminology of \citet{pankratev2026powerful}.
    A \emph{cluster} is a cross-sectional entity.
    A \emph{cell} is a cluster-by-time-period combination and serves as the unit of randomization.
    An \emph{individual observation} is a unit within a cell.
    \emph{Randomization-unit-level variance} is the spread in the average outcome across cells, driven by shocks common to all observations within a cell rather than specific to individual units.
    We use the former term interchangeably with \emph{cell-level variation}, \emph{macro (outcome) variation}, and \emph{macro variance}.
    This contrasts with idiosyncratic variation across observations within the same cell.
    In the data-generating process of Section~\ref{sec:switchback}, idiosyncratic outcome variation is modeled by the residual term $\epsilon_{i,j,h}$, and we use \emph{idiosyncratic variation} and \emph{residual variance} interchangeably.
}
Standard CUPED uses only a single lag of that cell average.
Enriching the covariate adjustment with additional pre-period lags and structured fixed effects for time and space is a simple extension that requires no custom feature construction.
Supplementary lags add information when outcome dynamics persist beyond the immediately preceding period.
Fixed effects for time and space account for stable cluster- and period-level differences that lagged values alone leave unexplained.
Together, these extensions more directly adjust for predictable cell-level variation and thereby increase statistical power in switchback experiments.

In this paper we develop this enriched adjustment and propose it as a readily deployable extension of standard CUPED for switchback experiments and other clustered designs.\footnote{
    Although we develop and evaluate our framework for switchback experiments, it broadly applies to other randomized experimental designs.
}
We motivate the design of our extended CUPED framework using the theoretical variance of the switchback treatment-effect estimator established by \citet{pankratev2026powerful}, including closed-form ceilings on how much variance cell-level adjustment can remove and how those ceilings depend on the composition of macro outcome variance.
We validate it in synthetic simulations spanning cluster- and time-shock-dominated, balanced, and cluster--time interaction-shock-dominated regimes, and in publicly available New York City taxi trip records, where a synthetic treatment is injected to evaluate accuracy against a known effect.
We show that adding multiple pre-period lags and hour-of-day structure in the prediction model---cyclic encodings plus low-dimensional hour fixed-effect dummies---consistently reduces uncertainty in the treatment-effect estimate, whereas incorporating high-dimensional spatial fixed-effect dummies into the prediction model can increase model complexity without a matching improvement in variance reduction.
We further show that fitting the enriched covariate set on the training data can produce spurious apparent efficiency gains that disappear under evaluation on held-out data, so cross-fitting is not an optional refinement but an essential part of the extension.

The remainder of the paper is organized as follows.
Section~\ref{sec:literature} reviews related work.
Section~\ref{sec:switchback} develops the multilevel switchback DGP, the variance of the treatment-effect estimator, composition-dependent variance-reduction ceilings, and the enriched CUPED specification.
Section~\ref{sec:implementation} presents a Monte Carlo simulation study and an empirical validation on NYC taxi trip data with injected treatment.
The simulation study compares basic and enriched CUPED across variance-composition regimes; ablates lag depth and fixed-effect choices in the pre-period prediction model; demonstrates the role of cross-fitting; and quantifies when multilag enrichment yields the largest absolute standard-error gains.
Section~\ref{sec:conclusion} concludes; supplementary tables underlying the figures are in Appendix~\ref{app:tables}.

\section{Literature Review}
\label{sec:literature}

CUPED was introduced by \citet{deng2013improving} as a model-free method for reducing uncertainty in treatment-effect estimates using pre-experiment data, drawing on classical control-variate ideas from Monte Carlo simulation \citep{owen2013monte}.
A substantial literature interprets and extends the technique.
\citet{deng2023augmentation} present CUPED as a general augmentation framework, clarifying its relationship to regression adjustment and extending it to ratio and percentile metrics, while also showing that in-experiment data can yield larger reductions than pre-experiment data alone.
The connection between covariate adjustment and the efficiency of experimental estimates is well established in the statistics literature \citep{freedman2008regression, lin2013agnostic}, where regression adjustment is known to improve precision when applied correctly.

A parallel line of work replaces the linear covariate with the prediction of a machine-learning model.
\citet{poyarkov2016boosted} use boosted decision trees to predict user outcomes from pre-experiment features and employ the prediction as an adjustment covariate, an approach now widely adopted in industry under the name CUPAC.
\citet{lin2024variance} extend the idea further by combining pre-experiment and in-experiment covariates, achieving greater uncertainty reduction than pre-experiment data alone permits.
The use of cross-fitting to remove the bias introduced by estimating a flexible prediction model is standard in the double machine-learning literature \citep{chernozhukov2018double}, and we adopt it here to ensure that realized efficiency gains are not inflated by in-sample overfitting.

Our analysis of variance reduction in clustered designs draws on the literature on cluster-randomized trials, which establishes that cluster size imbalance inflates uncertainty in the treatment-effect estimate, with the inflation growing with the dispersion of cluster sizes \citep{eldridge2006sample, moulton1986random}.
We rely specifically on the analytical framework for switchback experiments developed by \citet{pankratev2026powerful}, which separates outcome variation into spatial, temporal, interaction, and residual components and shows that these components respond to imbalance in qualitatively different ways.
The design properties of switchback experiments, including the limits of fixed assignment designs such as pairing and mirroring, are studied by \citet{bojinov2023design}.
\citet{pankratev2026design} evaluate CUPED, CUPAC, and doubly robust estimators in switchback designs across a range of settings, mapping when each method yields the largest efficiency gains while treating the covariate model as given.
Complementary to that comparative study, \citet{pankratev2027poweroptimal} ask how a CUPAC model should be trained and adjusted when randomization occurs at the cell level, deriving training and analysis procedures matched to the variance structure of switchback experiments.
The present paper addresses a distinct but related question: given a standard prediction objective, which covariates should enter the adjustment, and how do additional lags and fixed effects interact in reaching the cell-level outcome variation that drives statistical power?

\section{Theoretical Framework}
\label{sec:switchback}

In switchback experiments, treatment is assigned at the cell level, so the value of covariate adjustment depends on which components of outcome variation propagate into the treatment-effect estimate.
We build on the multilevel outcome model and variance decomposition of \citet{pankratev2026powerful} and use it to motivate an extended CUPED specification for switchback panels.

We model the outcome for individual unit $i$ in cell $(j, h)$---spatial cluster $j$, time period $h$---as
\begin{equation}
\label{eq:dgp}
Y_{i,j,h} = \mu + \tau W_{j,h} + \alpha_j + \gamma_h + \delta_{j,h} + \epsilon_{i,j,h},
\end{equation}
where $W_{j,h} \in \{0, 1\}$ is the cell-level treatment indicator, $\alpha_j$ is the spatial cluster main effect, $\gamma_h$ is the temporal period main effect, $\delta_{j,h}$ is the cluster-by-time interaction, and $\epsilon_{i,j,h}$ is within-cell idiosyncratic noise, with all error components mutually independent and mean zero.
We call $\vmac = \Var(\alpha_j) + \Var(\gamma_h) + \Var(\delta_{j,h})$ the \emph{macro} variance---shocks shared by every unit in a cluster, a period, or their intersection---and $\vres = \Var(\epsilon_{i,j,h})$ the \emph{residual} variance, idiosyncratic to each unit.

The individual-level OLS treatment-effect estimator regresses the outcome on treatment and pre-experiment covariates,
\begin{equation}
\label{eq:ols}
Y_{i,j,h} = \tau W_{j,h} + \beta X_{i,j,h} + \nu_{i,j,h},
\end{equation}
where $X_{i,j,h}$ collects the covariates used for adjustment and $\nu_{i,j,h}$ absorbs all remaining variation, including the intercept, the macro shocks, and the residual noise.
Ordinary least squares (OLS) yields an estimate $\hat\tau$ of the treatment effect.

Let $J$ denote the number of clusters, $H$ the number of periods, $\nbar$ the mean cell size, and $\cv$ the coefficient of variation of cell sizes.\footnote{
    Writing $n_{j,h}$ for the number of observations in cell $(j,h)$, $\cv^2 = \frac{1}{JH\nbar^2}\sum_{j=1}^{J}\sum_{h=1}^{H}(n_{j,h}-\nbar)^2$.
}
The asymptotic variance of the individual-level OLS estimator for $\tau$ is then
\begin{equation}
\label{eq:variance}
\Var(\hat\tau) \approx \frac{4}{J H}
\left[ \frac{\vres}{\nbar} + \vmac\left( \frac{1}{\nbar} + 1 + \cv^2 \right) \right].
\end{equation}
Equation~\eqref{eq:variance} separates the contribution of idiosyncratic (residual) variance, attenuated by mean cell density through the factor $1/\nbar$, from the contribution of cell-level variance, which is amplified by cluster size imbalance through $\cv^2$.
In typical switchback panels with moderate cell density and meaningful size imbalance, the macro term dominates the variance of the ATE estimator.
Cell-level outcome variation outweighs idiosyncratic residual variation in determining $\Var(\hat\tau)$.
Covariates that predict cluster-level, time-level, or cluster-by-time variation are therefore disproportionately valuable relative to their raw explanatory power.

When no covariate adjustment is used, OLS is run on the raw outcome $Y_{i,j,h}$ against treatment alone.
In that case, $\hat\tau$ can absorb variation in pre-period outcomes that correlates with treatment assignment, inflating the variance of the treatment-effect estimate.
Covariate adjustment---whether through CUPED or CUPAC---alleviates this problem by including a pre-experiment covariate $X_{i,j,h}$ in \eqref{eq:ols}, or equivalently by adjusting the outcome with its predicted value.\footnote{
    Covariate adjustment reduces variance to the extent that the chosen model accurately estimates $X_{i,j,h}$.
    With adjustment coefficient $\beta$, the equivalent outcome adjustment is
    \[
    Y_{i,j,h}^{\text{cuped}} = Y_{i,j,h} - \beta \left( X_{j,h} - \E[X_{j,h}] \right),
    \]
    which corresponds to including $X_{i,j,h}$ in \eqref{eq:ols}, so downstream variance computation and hypothesis testing remain identical to those of standard CUPED.
}
With standard CUPED, $X_{i,j,h}$ is a single lag of the pre-period outcome, typically constructed at the cell level in switchback settings.
With CUPAC, $X_{i,j,h}$ is instead the prediction of a flexible model trained on rich pre-experiment features \citep{poyarkov2016boosted}.
Under the usual switchback randomization scheme, in which treatment-assignment draws are independent across time and space, this adjustment does not bias average treatment effect (ATE) estimates by default.
Because that prediction model may include features that correlate with current-period treatment assignment, CUPAC can bias ATE estimates in settings where such correlations arise.
Because switchback panels are typically unbalanced, pre-period covariates are almost always aggregated to the cell, so $X_{i,j,h} = X_{j,h}$ is constant within a cell.

Our extended CUPED specification enriches $X_{j,h}$ with three elements suggested by this structure.
The first is multiple pre-period lags of the cell-level outcome, which capture persistent dynamics that a single lag misses.
The second is hour-of-day structure in the pre-period prediction model: cyclic encodings (always included in our simulations) and, optionally, low-dimensional hour fixed-effect dummies that capture period-level deviations beyond the smooth encoding.
The third is cross-fitted ridge regression, which fits the enriched feature set without in-sample overfitting.
In our baseline production specification, we omit spatial fixed-effect dummies from the pre-period prediction model because their inclusion increases the variance of the treatment-effect estimator, as shown in the next section.
We train the pre-period prediction model $g(\cdot)$ on cell-level features and apply $K$-fold cross-fitting so that each cell's out-of-fold covariate $X_{j,h}$ is fit without using that cell's own outcome.
In applied settings, folds should respect the randomization structure---for example, partitioned by cluster or by week---so that training data for a cell's prediction are disjoint from the cell itself.
In our Monte Carlo simulation (Section~\ref{sec:sim-study}), we instead use five-fold cross-validation with cell rows randomly assigned to folds; because treatment is independently randomized in each cell, this split is leakage-safe despite not grouping by cluster or week.
In the NYC validation (Section~\ref{sec:nyc}), folds are grouped by pickup-zone-by-hour cell.
Downstream analysis then proceeds as in standard CUPED, with $X_{i,j,h}=X_{j,h}$ constant within each cell.

When pre-period covariates are aggregated to the randomization unit, cell-level CUPED can target the macro shocks in \eqref{eq:dgp} but not the within-cell residual $\epsilon_{i,j,h}$.
The adjusted estimator therefore inherits at least the first term in \eqref{eq:variance}, while any variance reduction comes from shrinking the second.
Writing $\kappa = 1/\nbar + 1 + \cv^2$ and expressing $\vres$ and $\vmac$ as shares $\Sres$ and $\Smacro$ of total outcome variance, the unadjusted variance is proportional to $\Sres/\nbar + \Smacro\,\kappa$, and the smallest variance a cell-level adjustment can reach---with every recoverable macro component removed---is proportional to $\Sres/\nbar$ alone.
The implied optimistic ceiling on confidence-interval reduction, obtained by eliminating the entire macro term in \eqref{eq:variance}, is therefore
\begin{equation}
\label{eq:ceiling}
\VRmax = 1 - \sqrt{\frac{\Sres/\nbar}{\Sres/\nbar + \Smacro\,\kappa}},
\end{equation}
where $\VRmax = 100(1 - \mathrm{SE}_{\min}/\mathrm{SE}_{\text{raw}})$ in the notation of Section~\ref{sec:implementation}.
When a positive share of $\Sint$ is transient and unpredictable across the pre/post boundary, enrichment can remove $\Scl$, $\Stime$, and the recurring share $r$ of $\Sint$, but the transient part $(1-r)\Sint$ remains in the macro term.
The smallest variance enrichment can then attain is proportional to $\Sres/\nbar + (1-r)\Sint\,\kappa$, giving the tighter, composition-specific ceiling
\begin{equation}
\label{eq:ceiling-attainable}
\VRatt = 1 - \sqrt{\frac{\Sres/\nbar + (1-r)\Sint\,\kappa}{\Sres/\nbar + \Smacro\,\kappa}}.
\end{equation}
Equation~\eqref{eq:ceiling} is an upper bound under full macro removal; \eqref{eq:ceiling-attainable} is the bound we use to evaluate basic and enriched CUPED when $\Smacro$ is fixed but its composition across $\Scl$, $\Stime$, and $\Sint$ varies.

Although we develop the framework for switchback panels, the same enrichment applies whenever treatment is randomized at the cluster level and pre-period covariates are aggregated to the randomization unit.
Cluster-randomized trials are the leading example: multiple lags and low-dimensional temporal fixed effects target persistent cluster-level and period-level variation in the same way, and cross-fitting remains essential whenever the prediction model is fit on the same data used for inference.
The feature set and model class are not tied to switchbacks.
The same cross-fitted ridge template carries over unchanged, but practitioners may substitute alternative covariates---for example, longer lag windows, unit-level pre-period summaries when the randomization unit is an individual, or flexible learners when richer pre-experiment attributes are available---provided the chosen features respect the randomization structure and are evaluated out of sample.

\section{Empirical Evaluation}
\label{sec:implementation}

This section evaluates the extended CUPED framework in a Monte Carlo simulation study and in an empirical validation on publicly available NYC taxi trip data.

\subsection{Simulation Study}
\label{sec:sim-study}

In this subsection we present the design and results of the simulation study that tests and validates our enriched CUPED framework.
It is designed to separate two channels through which enrichment improves on basic CUPED: fixed-effect absorption of spatial and temporal main effects, and lag-depth averaging of the cluster-by-time interaction.

\subsubsection{Simulation Design}
\label{sec:simdesign}
We simulate switchback panels from the multilevel decomposition in \eqref{eq:dgp}, drawing spatial, temporal, interaction, and residual components as mutually independent mean-zero Gaussian shocks and drawing cell sizes from a lognormal-Poisson mixture to represent imbalanced panels.
Every regime shares the same design geometry: $J=200$ spatial units, $H=24$ hourly periods, mean cell size $\nbar = 180$, cluster-size coefficient of variation $\cv=1.5$, and three weeks of pre-experiment history, with cluster-robust standard errors throughout.
Pre-period cell counts are $n^{\mathrm{pre}}_{j,h}\sim\mathrm{Poisson}(\rho_{\mathrm{pre}}\bar n)$ with $\rho_{\mathrm{pre}}=0.02$.

We decompose the interaction term $\delta_{j,h,t}$ into a sum of a recurring component $\delta^{\mathrm{rec}}_{j,h}$ and a transient component $\delta^{\mathrm{trans}}_{j,h,t}$.
The recurring component $\delta^{\mathrm{rec}}_{j,h}$ is a fixed cluster-by-hour baseline that is constant across pre-experiment weeks, while the transient component $\delta^{\mathrm{trans}}_{j,h,t}$ is a week-specific shock that evolves across weeks.
We set $\Var(\delta^{\mathrm{rec}}_{j,h})/\Var(\delta_{j,h,t})=0.85$, so the remaining $15\%$ of interaction variance is carried by $\delta^{\mathrm{trans}}_{j,h,t}$, which we model as a stationary AR(3) process across weeks.
With this structure, the $k$-th pre-period covariate is the cell mean $\bar Y_{j,h}^{(-k)}$.
Under these conditions, additional lags carry information beyond a single pre-period value, and averaging them reduces measurement noise.

We consider three regimes that differ in the composition of outcome variance: whether $\vmac$ is concentrated in cluster- and time-period shocks ($\Scl+\Stime$) or in cluster--time interaction shocks ($\Sint$).
We hold $\Smacro=0.35$ and $\Sres=0.65$ fixed and label them as follows:
\begin{itemize}
    \item cluster- and time-shock-dominated ($\Scl=\Stime=0.15$, $\Sint=0.05$)
    \item balanced ($\Scl=\Stime=0.10$, $\Sint=0.15$)
    \item cluster--time interaction-shock-dominated ($\Scl=\Stime=0.05$, $\Sint=0.25$)
\end{itemize}

Each configuration is evaluated over $300$ independent replications with cluster-robust standard errors.

Plugging the shared simulation parameters ($\nbar=180$, $\cv=1.5$, $\Sres=0.65$, $\Smacro=0.35$) into \eqref{eq:variance} gives $\kappa=3.26$ and an unadjusted variance proportional to $\Sres/\nbar + \Smacro\,\kappa \approx 1.14$.
Because $\Sres$ and $\Smacro$ are held fixed across the three regimes, this unadjusted variance and the optimistic full macro-removal ceiling $\VRmax \approx 94\%$ from \eqref{eq:ceiling} are identical in each regime.
The attainable ceiling $\VRatt$ from \eqref{eq:ceiling-attainable} does vary with $\Sint$ (and with $r=0.85$): approximately $84\%$ in the cluster- and time-shock-dominated regime, $74\%$ in the balanced regime, and $67\%$ in the cluster--time interaction-shock-dominated regime.
In all three cases the macro term accounts for roughly $99.7\%$ of the formula's variance weight, so the scope for cell-level CUPED is governed almost entirely by the second term in \eqref{eq:variance}.

For enriched CUPED, we fit the pre-period prediction model at the cell level, using pre-period cell aggregates as features.\footnote{
    Because treatment is randomized at the cell and all pre-period features are cell aggregates, $g(\cdot)$ is fit on the $J \times H$ cell panel and the resulting covariate is constant within each cell, consistent with \eqref{eq:ols} and standard switchback CUPED practice.
}
The enriched feature set in $g(\cdot)$ always includes the cell mean outcome from each pre-experiment week (one or three lags), the per-cluster mean pooled over those weeks, and cyclic hour-of-day encodings (sine and cosine of hour).
The feature-set ablation toggles one-hot \emph{fixed-effect dummies} for cluster and hour on top of that base set; it does not remove the encodings.

In our analysis, we compare three estimators throughout the simulation study: an unadjusted difference in means, basic CUPED with a single lagged cell mean, and enriched CUPED with a cross-fitted ridge pre-period model.
Enriched CUPED is evaluated under several feature sets that vary lag depth and whether cluster and hour fixed-effect \emph{dummies} enter $g(\cdot)$.
All adjusted estimators apply the standard CUPED adjustment---centering the control variate, subtracting $\hat\theta(X_{j,h}-\bar X)$ from the outcome, and regressing the adjusted outcome on treatment---with cluster-robust standard errors; we do not add cluster or hour fixed effects in that final regression.

\subsubsection{DGP Specification and Variance Reduction Efficiency}
\label{sec:emp-composition}

This subsection asks how the composition of outcome variance shapes variance-reduction efficiency along the axis from cluster/time shocks to cluster--time interaction shocks---the cluster- and time-shock-dominated, balanced, and cluster--time interaction-shock-dominated regimes defined in Section~\ref{sec:simdesign}.
We first characterize the variance reduction delivered by basic CUPED, which adjusts on a single lagged cell mean, relative to the unadjusted estimator in each regime.
We then compare that baseline to enriched CUPED and measure how much additional variance reduction a richer pre-period feature set provides.

\begin{figure}[H]
\centering
\includegraphics[width=0.7\textwidth]{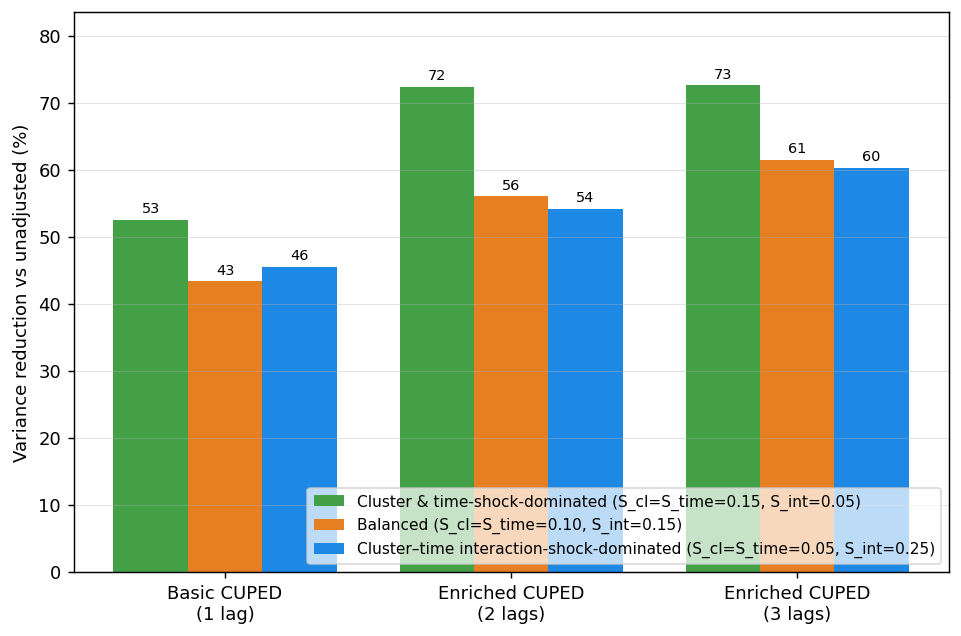}
\caption{Variance reduction by regime.}
\label{fig:headline}
\end{figure}

Figure~\ref{fig:headline} reports these comparisons by regime, along with the incremental gap in percentage points.\footnote{
    Variance reduction is $100 \times (1 - \mathrm{SE}_{\text{adjusted}} / \mathrm{SE}_{\text{unadjusted}})$; the gap is enriched minus basic CUPED.
}

Holding $\Smacro = 0.35$ and $\Sres = 0.65$ fixed, basic CUPED uses one pre-period lag throughout.
Enriched CUPED uses three pre-period weekly cell means, cyclic hour-of-day encodings, and both cluster- and hour-level fixed-effect dummies in the pre-period prediction model.\footnote{
    
    Section~\ref{sec:emp-specification} shows that hour fixed-effect dummies alone in $g(\cdot)$ perform best (on top of the always-included encodings).
    We use the both-dummy specification here for comparability across simulation experiments, where the gap to the hour-dummy-only variant is one to two percentage points in variance reduction.
    
}
Across the three regimes, enriched CUPED consistently improves on basic CUPED in terms of the efficiency of variance reduction.
The incremental gain in variance reduction ranges from $+14.8$ to $+20.1$ percentage points, largest in the cluster- and time-shock-dominated regime and smallest in the cluster--time interaction-shock-dominated regime.
Absolute variance reduction under enrichment spans roughly $60\%$--$73\%$, compared with $43\%$--$53\%$ under basic CUPED.

The increase in variance-reduction efficiency from CUPED enrichment falls monotonically from 20.1 to 14.8 as $\Sint$ rises from $0.05$ to $0.25$.
This pattern is intuitive: enrichment adds the most over basic CUPED when macro variance sits in cluster- and time-period shocks that fixed-effect dummies in $g(\cdot)$ can absorb, and the incremental return narrows as a larger share of $\vmac$ shifts into $\Sint$, where a single pre-period lag already recovers much of the recurring component.
In the cluster- and time-shock-dominated regime, hour and cluster fixed-effect dummies in the prediction model absorb most of the cluster- and time-period variation that a single noisy lag captures only partially, producing the largest gap.
As the cluster--time interaction share rises, these dummies can no longer address interaction shocks, and further gains must come from lag-depth averaging.
Because the recurring interaction share is high ($0.85$), a single lag already captures much of that component, so the marginal benefit of three lags narrows even though absolute variance reduction remains large ($60.3\%$ in the cluster--time interaction-shock-dominated regime).
Enrichment is therefore robust across the three regimes, with the largest incremental advantage over basic CUPED in the cluster- and time-shock-dominated regime.

Relative to the attainable ceiling $\VRatt$ in \eqref{eq:ceiling-attainable} with $r=0.85$, basic CUPED captures $58\%$--$68\%$ of the bound in the three regimes ($52.5\%$, $43.4\%$, and $45.5\%$ realized against $\VRatt \approx 84\%$, $74\%$, and $67\%$).
Enriched CUPED closes most of the remaining gap, reaching $83\%$--$90\%$ of the same bound ($72.6\%$, $61.5\%$, and $60.3\%$ realized).
The cluster--time interaction-shock-dominated regime has the lowest $\VRatt$ because a larger $\Sint$ share leaves more transient interaction in the macro term of \eqref{eq:variance}, and enrichment approaches that lower bound most closely.

\subsubsection{Model Specification and Variance Reduction Efficiency}
\label{sec:emp-specification}
We next examine two specification choices in the pre-period prediction model $g(\cdot)$: which features to include, and whether to cross-fit.

We first evaluate eight enriched CUPED configurations on the cluster--time interaction-shock-dominated regime, cross-cutting two lag specifications---one pre-period weekly cell mean versus three---with the inclusion of cluster (spatial) and hour (temporal) fixed-effect \emph{dummies} in the ridge prediction model.
All eight configurations retain the cyclic hour-of-day encodings and per-cluster pre-period mean.
We use this regime because it assigns the largest share of $\vmac$ to $\Sint$, the component that fixed effects cannot remove.
All estimators share the same downstream step: cluster-robust OLS of the adjusted outcome on treatment only.
Spatial and temporal main effects are therefore captured only to the extent that the pre-period prediction model $g(\cdot)$---and the per-cluster pre-period mean already in the base feature set---partially predict them in the control variate.
Table~\ref{tab:featureset} reports the mean percentage of variance reduction (VR) relative to the unadjusted estimator over replications.

\begin{table}[H]
\centering
\caption{Variance reduction by feature set.}
\label{tab:featureset}
\begin{tabular}{lcc}
\toprule
Prediction Model Features & 1 Lag (VR) & 3 Lags (VR) \\
\midrule
Basic CUPED (production baseline) & 45.2\% & --- \\
Enriched CUPED: lags + encodings (no FE dummies) & 42.6\% & 57.5\% \\
Enriched CUPED: lags + encodings + hour FE dummies & 44.7\% & 61.8\% \\
Enriched CUPED: lags + encodings + cluster FE dummies & 38.6\% & 55.1\% \\
Enriched CUPED: lags + encodings + both FE dummies & 41.8\% & 59.9\% \\
\bottomrule
\end{tabular}
\end{table}

The no-dummy configurations isolate the effect of additional weekly lags together with the always-included hour encodings and per-cluster pre-period mean.
With one pre-period cell mean, this base enriched specification reaches $42.6\%$ variance reduction, slightly below the $45.2\%$ production basic CUPED baseline, because ridge regression penalizes large coefficient magnitudes and thereby attenuates the single-lag adjustment relative to unpenalized basic CUPED.
Adding the second and third weekly lags raises variance reduction to $57.5\%$, a gain of $14.9$ percentage points, so multivariate lag enrichment improves efficiency even without fixed-effect dummies in $g(\cdot)$.

Adding fixed-effect dummies to these lag specifications reveals a specific and asymmetric role for the two dummy types, rather than a blanket ``omit fixed effects'' rule: including the low-dimensional hour fixed-effect dummies in the pre-experiment prediction model consistently decreases estimator variance, while including the high-dimensional cluster fixed-effect dummies consistently increases it, at both lag specifications.
The best-performing configuration at either lag depth is \emph{lags + encodings + hour FE dummies}, reaching $44.7\%$ with one pre-period weekly cell mean (statistically indistinguishable from the $45.2\%$ production basic CUPED baseline) and $61.8\%$ with three.
Adding cluster fixed-effect dummies on top of the hour dummies degrades performance in both specifications ($44.7\%\to41.8\%$ with one lag and $61.8\%\to59.9\%$ with three), and the cluster-dummy-only configuration is the single worst option under either ($38.6\%$ and $55.1\%$).

The mechanism driving these results is the dimensionality asymmetry between space and time fixed-effect dummies.
Hour fixed-effect dummies add only $H=24$ categories, so including them in the cross-fitted ridge regression costs almost nothing in estimation variance while letting the model capture hour-specific deviations beyond the smooth cyclic encodings (which are always present).
Cluster fixed-effect dummies add $J=200$ categories, a substantial fraction of the roughly $4{,}800$ spatial-by-hour cells the regression is fit on.
Even under ridge shrinkage and cross-fitting, estimating $200$ additional coefficients materially inflates the out-of-fold variance of $X_{j,h}$, and this variance cost is not offset by any corresponding gain in predictive signal: the per-cluster pre-period mean in the base feature set already summarizes cluster-level variation, so the cluster dummies are largely redundant for targeting $\Scl$.
Because the final regression does not include its own cluster fixed effects, there is nothing downstream that makes high-dimensional cluster dummies in $g(\cdot)$ costless.

We conclude that the pre-experiment prediction model should include hour fixed-effect dummies but exclude cluster fixed-effect dummies.
The remaining simulation results use the fully enriched specification (both dummy types in $g(\cdot)$, with encodings always included) for comparability across experiments.
The hour-dummy-only variant is the preferred production configuration and differs by only one to two percentage points in variance reduction here.

We next turn to the second specification choice, cross-fitting, and make concrete the finite-sample failure mode it guards against, namely in-sample overfitting.
In the Monte Carlo study, $g(\cdot)$ is fit with five-fold cross-validation on the cell panel using random fold assignment, as described in Section~\ref{sec:switchback}.
To illustrate this failure mode, we conduct a simulation where we construct the cell-level control variate using an unregularized ordinary least squares linear regression model.
We then progressively add pure-noise features (independent, identically distributed standard Gaussian variables drawn independently of the outcome and of any true covariates) to the model, and evaluate the impact on the cell-level $R^2$ of the fitted control variate.
We contrast in-sample $R^2$ with cross-fitted $R^2$ as the number of pure-noise features increases (Figure~\ref{fig:q3}; See Appendix~\ref{app:tables}, Table~\ref{tab:q3}, for full results).

As the pure-noise features accumulate, in-sample $R^2$ rises spuriously from $36.9\%$ to $51.5\%$, while cross-fitted $R^2$ declines from $19.9\%$ to $11.0\%$.
This decline reflects the degrees-of-freedom cost of fitting parameters that carry no predictive signal.
Cross-fitting is therefore a core requirement for the enriched covariate set: it prevents spurious efficiency gains in finite samples and allows a rich feature set without manual feature curation.

\begin{figure}[H]
\centering
\includegraphics[width=0.8\textwidth]{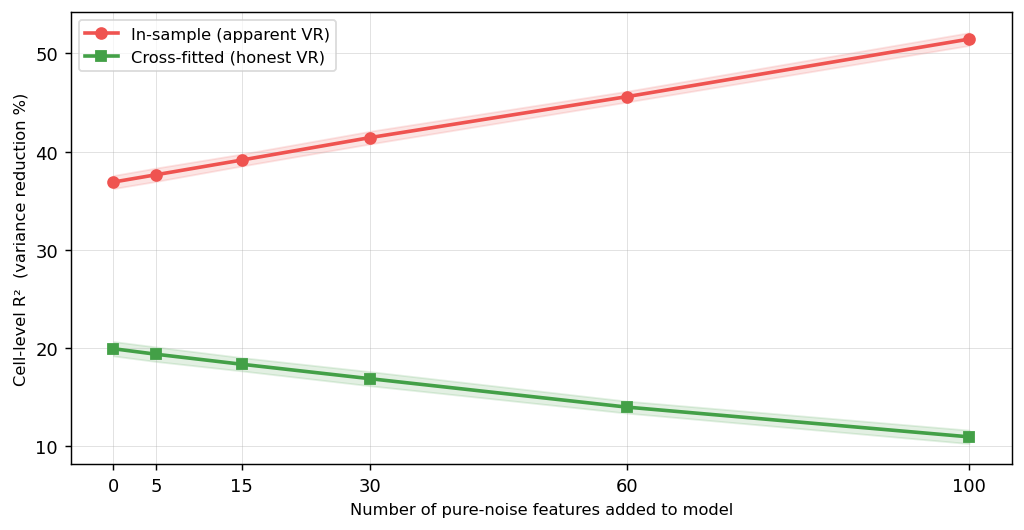}
\caption{Cell-level $R^2$ of the control variate under in-sample and cross-fitted evaluation across number of noise features.}
\label{fig:q3}
\end{figure}

\subsubsection{When Enrichment Matters Most}
\label{sec:emp-q4}
We quantify when enrichment yields the largest absolute gains in standard-error units and isolate the role of recurring interaction signal.

\begin{figure}[H]
\centering
\includegraphics[width=\textwidth]{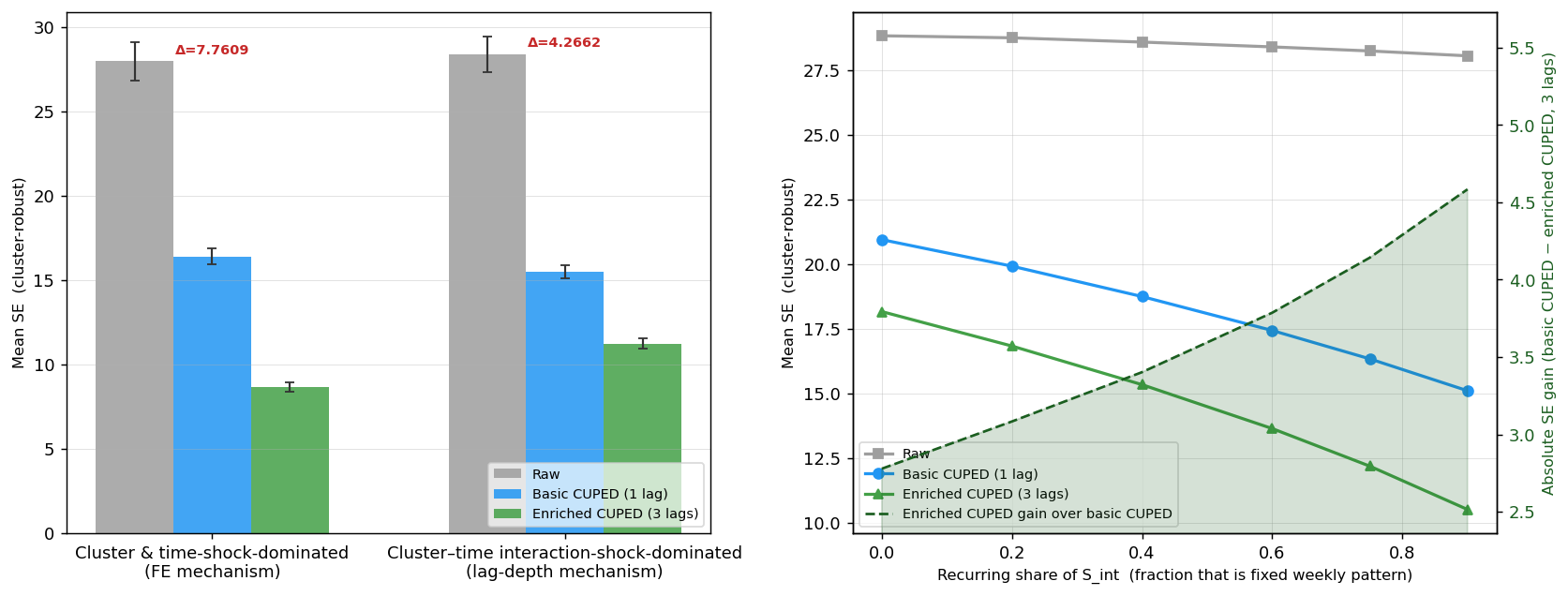}
\caption{Standard errors by regime and recurring interaction share.
In the left panel, annotated $\Delta$ values are $\SE(\text{basic CUPED})-\SE(\text{enriched CUPED, 3 lags})$.}
\label{fig:q4}
\end{figure}

These simulations are run separately from the headline comparison of Section~\ref{sec:emp-composition}, with $150$ independent replications per configuration.
Throughout, basic CUPED adjusts on a single pre-period cell mean and enriched CUPED fits a cross-fitted ridge pre-period model on three pre-period weekly cell means, the cluster pre-period mean, cyclic hour-of-day encodings, and both cluster- and hour-level fixed-effect dummies in $g(\cdot)$---the same enriched specification used in Section~\ref{sec:emp-composition}.
The left panel of Figure~\ref{fig:q4} re-estimates absolute standard errors in the cluster- and time-shock-dominated and cluster--time interaction-shock-dominated regimes defined in Section~\ref{sec:simdesign}; it does not extend $\Sint$ beyond the $0.05$--$0.25$ range of that comparison.
The right panel of Figure~\ref{fig:q4} holds the interaction-shock-dominated regime fixed at $\Sint=0.25$ and sweeps only the recurring share of $\Sint$ from zero to $0.9$, with $150$ replications at each of six recurring-share values (See Appendix~\ref{app:tables}, Table~\ref{tab:q4b}, for full results).

Figure~\ref{fig:q4} (left panel) compares standard errors in the cluster- and time-shock-dominated and cluster--time interaction-shock-dominated regimes.
The absolute standard-error gain of three-lag enriched CUPED over basic CUPED is $7.76$ in the former and $4.27$ in the latter.

The right panel of Figure~\ref{fig:q4} isolates the lag-depth mechanism by holding all other parameters constant and varying only the recurring share of $\Sint$ in the cluster--time interaction-shock-dominated regime.
Because the recurring part of the interaction is a stable cluster-by-time-of-week baseline, every week's pre-period observation represents an independent, noisy measurement of the exact same underlying pattern.
As the recurring share increases from zero to $0.9$, the underlying signal becomes a larger fraction of the total interaction variance, and the second and third lags provide increasingly accurate independent observations of this stable signature.
By averaging these multiple lags, the regression model effectively drives down the measurement noise.
As a result, the absolute advantage of three-lag enriched CUPED over basic CUPED grows monotonically, from $2.78$ to $4.58$ standard-error units.

The advantage remains positive ($2.78$ units) even at a zero recurring share because the AR(3) transient interaction retains serial memory, so additional lags remain partially informative even without a recurring weekly pattern.
Under an AR(1) transient, by contrast, a single lag would be sufficient and additional lags would add little.
These results indicate that multivariate lag enrichment is most valuable when the cluster-by-time interaction is persistent, whether through a recurring weekly pattern or longer serial memory---the macro component that fixed assignment designs cannot balance and that post-experiment covariate enrichment is designed to address.

\subsection{Empirical Validation on NYC Taxi Trip Data}
\label{sec:nyc}

To assess whether the same patterns hold under realistic seasonality and imbalance, we apply the enriched CUPED framework to one month of publicly available New York City Taxi and Limousine Commission (TLC) yellow-taxi trip records (January 2024), using trip duration as the outcome and pickup-zone-by-hour as the cell.
Because no real experiment was run on this data, we inject a synthetic cell-level Bernoulli($0.5$) treatment with a known effect $\tau=-30$ seconds into one week of the month.
The three preceding, untreated weeks supply the lagged covariates $y_{\mathrm{lag}1}$--$y_{\mathrm{lag}3}$ that a real switchback analysis would use.
Cells with fewer than $10$ trips in any of the four weekly windows are dropped, since a sparse lag window cannot supply a reliable covariate.
This removes $52.9\%$ of candidate cells ($3{,}066 \to 1{,}443$), leaving an analysis panel of $676{,}426$ trips.

Decomposing variation in the raw cell means shows that this panel most closely resembles the cluster- and time-shock-dominated regime (Section~\ref{sec:simdesign}): $73.5\%$ spatial ($\Scl$), $10.5\%$ temporal ($\Stime$), and $15.9\%$ interaction ($\Sint$), so $\Scl+\Stime$ account for $84.0\%$ of macro outcome variance.

\begin{figure}[H]
\centering
\includegraphics[width=0.95\textwidth]{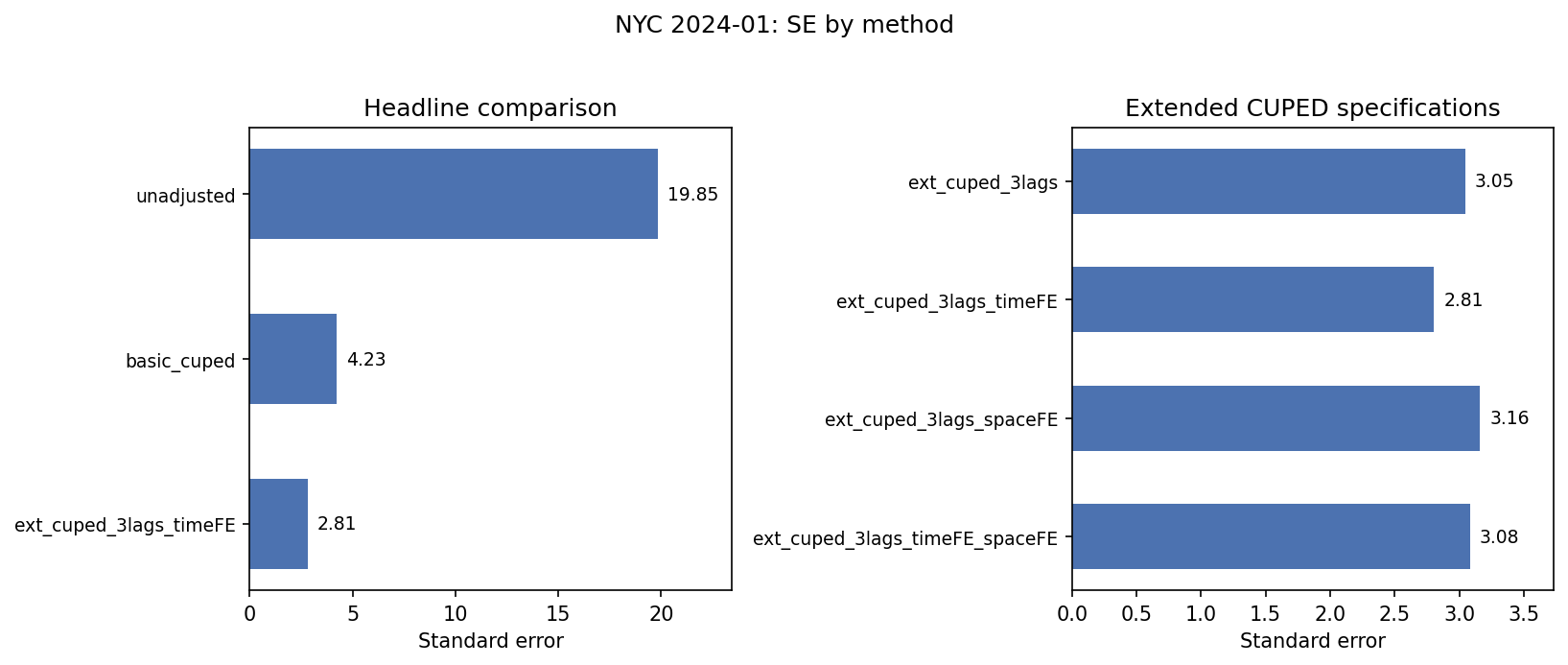}
\caption{Standard error by method (NYC TLC trips, January 2024).
}
\label{fig:nyc}
\end{figure}

We compare the unadjusted difference in means, basic CUPED using $y_{\mathrm{lag}1}$ as a scalar covariate (the production baseline), and four cross-fitted three-lag enriched CUPED configurations that vary which fixed-effect \emph{dummies} enter the prediction model $g(\cdot)$---none, temporal (hour-of-day and day-of-week dummies), spatial (pickup zone), and both---mirroring the feature-set comparison of Section~\ref{sec:emp-specification}.
All four enriched configurations retain cyclic hour-of-day encodings and the per-zone pre-period mean.
Because day-of-week varies within a pickup-zone-by-hour cell over the seven-day estimation window, the control variate here is fit at the trip level rather than the cell level, with cross-fitting folds grouped by cell so that no trip's own cell contributes to its out-of-fold prediction.
Figure~\ref{fig:nyc} summarizes the standard errors in two panels---a headline comparison of the unadjusted estimator, basic CUPED, and the best extended specification (left), and all four three-lag extended configurations (right) (See Appendix~\ref{app:tables}, Table~\ref{tab:nyc}, for full point estimates, standard errors, and variance reduction).

All adjusted point estimates fall within a few seconds of the injected $\tau=-30$s, while the unadjusted estimator's wide interval (SE $=19.85$; left panel of Figure~\ref{fig:nyc}) still covers it despite a visibly biased point estimate, consistent with the sampling noise a single untreated-versus-treated week induces.
Basic CUPED already removes $78.7\%$ of the unadjusted variance (SE $=4.23$).
Enriching to three cross-fitted lags recovers a further $6$ percentage points, reaching $85.9\%$ when the prediction model also includes hour- and day-of-week fixed-effect dummies on top of the encodings (SE $=2.81$).
The right panel of Figure~\ref{fig:nyc} shows the four extended specifications clustering between SE $=2.81$ and $3.16$.
This reproduces the asymmetry documented in Section~\ref{sec:emp-specification}: adding the high-cardinality zone fixed effect to the prediction model, alone or alongside the temporal one, degrades performance relative to omitting it, because the per-zone pre-period mean already captures much of the spatial main effect and the high-dimensional zone dummies only inflate out-of-fold prediction variance in $g(\cdot)$ without a corresponding benefit in the final cluster-robust treatment regression.
This is exactly what the composition of outcome variance in the real panel implies: with $84.0\%$ of macro outcome variance in $\Scl+\Stime$, most of the achievable gain is available to the fixed-effect mechanism rather than to lag-depth averaging on $\Sint$, mirroring the cluster- and time-shock-dominated regime of Section~\ref{sec:simdesign}.

For the NYC panel ($\nbar \approx 469$, $\cv=1.5$), \eqref{eq:ceiling} gives an optimistic full macro-removal ceiling of $\VRmax \approx 97\%$.
Substituting the panel's effective interaction share ($\Sint = 0.159 \times \Smacro \approx 0.056$ of total variance) and $r=0.85$ into \eqref{eq:ceiling-attainable}, with $\Sres=0.65$ and $\Smacro=0.35$ calibrated as in the simulation, yields $\VRatt \approx 84\%$, matching the cluster- and time-shock-dominated simulation regime.
Basic CUPED reaches $94\%$ of $\VRatt$ ($78.7\%$ realized), and enriched CUPED with temporal fixed-effect dummies essentially attains it ($85.9\%$ realized).

Because no real experiment was run on this data, the injected treatment effect---not the noise structure, the covariates, or the cluster-size imbalance---is synthetic.
The exercise validates that the enriched estimator and its cross-fitting procedure behave as the theory predicts under a real, non-Gaussian, seasonally structured outcome distribution, rather than constituting an independent test of an actual causal effect.
The results reported here come from a single treatment-assignment draw and a single calendar month, in contrast to the $300$-replication Monte Carlo estimates of Section~\ref{sec:sim-study}.
Extending this validation to multiple months and repeated treatment draws is left for future revisions of this analysis.

\section{Conclusion}
\label{sec:conclusion}

We develop an extended CUPED framework for switchback and clustered experiments in which multiple pre-period lags, cyclic hour-of-day encodings and low-dimensional hour fixed-effect dummies in the prediction model, and cross-fitted ridge regression enrich the cell-level covariate set while remaining straightforward to deploy.
The extension targets cell-level outcome variation---the component that cluster size imbalance inflates most severely---rather than idiosyncratic unit-level noise.

Simulation results show that enrichment yields large variance reductions across cluster- and time-shock-dominated, balanced, and cluster--time interaction-shock-dominated regimes, with the largest incremental gain over basic CUPED in the cluster- and time-shock-dominated regime.
Temporal hour fixed-effect dummies in the prediction model consistently help (on top of the encodings).
Spatial fixed-effect dummies in the prediction model can hurt by increasing model complexity without predictive benefit, because cluster-level variation is already summarized by the per-cluster pre-period mean in the base feature set.
Cross-fitting is essential: in-sample fitting of the enriched covariate set can report spurious efficiency gains that reverse on held-out data.

The closed-form ceilings in \eqref{eq:ceiling} and \eqref{eq:ceiling-attainable} organize these findings: when $\Smacro$ is fixed, shifting macro variance toward transient interaction lowers the attainable bound, and enriched CUPED approaches that bound far more closely than basic CUPED in both simulation and the NYC validation.

A validation on NYC taxi trip data with injected synthetic treatment reproduces the same method ordering and fixed-effect asymmetry found in simulation.
The framework is intended as a practical, automatable extension of production CUPED for switchback experiments rather than a replacement for flexible CUPAC models when rich individual-level features are available.

\appendix
\section{Supplementary Tables}
\label{app:tables}

The tables below report the numeric values underlying the figures in the main text.

\begin{table}[H]
\centering
\footnotesize
\caption{Cell-level $R^2$ of the control variate under in-sample and cross-fitted evaluation across number of noise features (Figure~\ref{fig:q3}).}
\label{tab:q3}
\begin{tabular}{@{}lcc@{}}
\toprule
Noise features & In-sample $R^2$ & Cross-fitted $R^2$ \\
\midrule
0   & 36.9\% & 19.9\% \\
5   & 37.7\% & 19.4\% \\
15  & 39.1\% & 18.4\% \\
30  & 41.4\% & 16.9\% \\
60  & 45.6\% & 14.0\% \\
100 & 51.5\% & 11.0\% \\
\bottomrule
\end{tabular}
\end{table}

\begin{table}[H]
\centering
\footnotesize
\setlength{\tabcolsep}{5pt}
\caption{Standard errors by recurring interaction share (right panel of Figure~\ref{fig:q4}). Recurring share is the fraction of $\Sint$ that persists across the pre/post boundary ($r$ in Section~\ref{sec:simdesign}).}
\label{tab:q4b}
\begin{tabular}{@{}ccccc@{}}
\toprule
Recurring share & Unadj.\ SE & Basic SE & Enriched SE & SE gain \\
of $\Sint$ & & (1 lag) & (3 lags) & \\
\midrule
0.00 & 28.82 & 20.95 & 18.17 & 2.78 \\
0.20 & 28.74 & 19.92 & 16.84 & 3.08 \\
0.40 & 28.57 & 18.75 & 15.35 & 3.40 \\
0.60 & 28.39 & 17.44 & 13.65 & 3.79 \\
0.75 & 28.23 & 16.35 & 12.20 & 4.14 \\
0.90 & 28.04 & 15.12 & 10.53 & 4.58 \\
\bottomrule
\end{tabular}
\end{table}

\begin{table}[H]
\centering
\footnotesize
\setlength{\tabcolsep}{5pt}
\vspace{0.75em}
\begin{tabular}{@{}L{0.48\textwidth}rrr@{}}
\toprule
Method & $\hat\tau$ & SE & VR \\
\midrule
Unadjusted & $-9.42$ & $19.85$ & --- \\
Basic CUPED & $-27.65$ & $4.23$ & $78.7\%$ \\
Enriched: lags + encodings (no FE dummies) & $-26.92$ & $3.05$ & $84.7\%$ \\
Enriched: lags + encodings + temporal FE dummies & $-26.70$ & $2.81$ & $85.9\%$ \\
Enriched: lags + encodings + spatial FE dummies & $-29.09$ & $3.16$ & $84.1\%$ \\
Enriched: lags + encodings + both FE dummies & $-28.76$ & $3.08$ & $84.5\%$ \\
\bottomrule
\end{tabular}
\caption{Treatment-effect estimates by method (NYC TLC trips, January 2024; Figure~\ref{fig:nyc}). Enriched configurations use three cross-fitted lags; temporal FE dummies include hour-of-day and day-of-week indicators in $g(\cdot)$.}
\label{tab:nyc}
\end{table}

\bibliography{references}

\end{document}